\documentclass[aoas,MSNbibl,nameyear,seceqn,dvips]{arximspdf}
\usepackage{algorithmic}
\usepackage{graphicx}

\doi{10.1214/11-AOAS476}
\volume{5}
\issue{2B}
\pubyear{2011}
\firstpage{1159}
\lastpage{1182}

\makeatletter
\renewcommand{\epsilon}{\varepsilon}

\renewcommand\Re{\operatorname{Re}}
\renewcommand\Im{\operatorname{Im}}
\def\Real{\mathbb{R}}
\def\E{\mathbb{E}}
\def\P{\mathbb{P}}
\newcommand{\smooth}{\operatorname{smooth}}
\newcommand{\argmax}{\operatorname{\arg\max}\limits}
\newcommand{\argmin}{\operatorname{\arg\min}\limits}
\def\imagedatabase{\mathcal{D}}
\def\bagofimages{\mathcal{B}}
\def\usefulvoxels{\mathcal{V}}
\newcommand{\df}{\operatorname{df}}
\newcommand{\eqref}[1]{(\ref{#1})}
\makeatother

\begin{document}
\begin{frontmatter}

\title{Encoding and decoding V1 fMRI responses to natural images with
sparse nonparametric models}
\runtitle{Encoding and decoding V1 fMRI}

\begin{aug}
\author[A]{\fnms{Vincent Q.} \snm{Vu}\corref{}\thanksref{a1}\ead[label=e1]{vqv@stat.cmu.edu}},
\author[B]{\fnms{Pradeep} \snm{Ravikumar}\thanksref{a2}\ead[label=e2]{pradeepr@cs.utexas.edu}},
\author[C]{\fnms{Thomas} \snm{Naselaris}\thanksref{a3}\ead[label=e3]{tnaselar@berkeley.edu}},
\author[D]{\fnms{Kendrick N.} \snm{Kay}\thanksref{a4}\ead[label=e4]{knk@stanford.edu}},
\author[E]{\fnms{Jack L.} \snm{Gallant}\thanksref{a5}\ead[label=e5]{gallant@berkeley.edu}}
\and
\author[F]{\fnms{Bin} \snm{Yu}\ead[label=e6]{binyu@stat.berkeley.edu}\thanksref{a6,a6b}}
\runauthor{V. Q. Vu et al.}
\affiliation{University of California, Berkeley}
\address[A]{V.~Q.~Vu\\
Department of Statistics\\
Carnegie Mellon University\\
Pittsburgh, Pennsylvania 15213\\
USA\\
\printead{e1}} 
\address[B]{P.~Ravikumar\\
Department of Computer Sciences\hspace*{6.5pt}\\
University of Texas, Austin\\
Austin, Texas 78712\\
USA\\
\printead{e2}}
\address[C]{T.~Naselaris\\
Helen Wills Neuroscience Institute\\
University of California, Berkeley\\
Berkeley, California 94720\\
USA\\
\printead{e3}}
\address[D]{K.~N.~Kay\\
Department of Psychology\\
Stanford University\\
Stanford, California 94305\hspace*{32.5pt}\\
USA\\
\printead{e4}}
\address[E]{J.~L.~Gallant\\
Helen Wills Neuroscience Institute\\
Department of Psychology\\
and\\
Vision Science Program\\
University of California, Berkeley\\
Berkeley, California 94720\\
USA\\
\printead{e5}}
\address[F]{B.~Yu\\
Department of Statistics\\
University of California, Berkeley\\
Berkeley, California 94720\\
USA\\
\printead{e6}}
\end{aug}
\thankstext{a1}{Supported by a National Science Foundation (NSF) VIGRE Graduate
Fellowship and NSF Postdoctoral Fellowship DMS-09-03120.}
\thankstext{a2}{Supported by NSF Grant IIS-1018426.}
\thankstext{a3}{Supported by a National Institutes of Health (NIH) postdoctoral award.}
\thankstext{a4}{Supported by a National Defense Science and Engineering Graduate
Fellowship.}
\thankstext{a5}{Supported by grants from the National Eye
Institute and NIH.}
\thankstext{a6}{Supported by NSF Grants DMS-09-07632 and
CCF-093970.}
\thankstext{a6b}{Senior first author, following the convention of
biology publications.}

\received{\smonth{10} \syear{2010}}
\revised{\smonth{4} \syear{2011}}

%
\begin{abstract}
Functional MRI (fMRI) has become the most common method for
investigating the human brain. However,
fMRI data present some complications for statistical analysis and
modeling. One recently
developed approach to these data focuses on estimation of computational
encoding models that
describe how stimuli are transformed into brain activity measured in
individual voxels.
Here we aim at building encoding models for fMRI signals recorded in
the primary visual cortex of the human brain. We use residual analyses
to reveal systematic nonlinearity across voxels not taken into account
by previous models. We then show how a sparse nonparametric method
[\textit{J. Roy. Statist. Soc. Ser. B}
\textbf{71} (2009b)
1009--1030] can be used together with correlation screening to
estimate nonlinear encoding models effectively. Our approach produces encoding
models that predict about 25\% more accurately than models estimated using
other methods [\textit{Nature}
\textbf{452}
(2008a)
352--355]. The estimated nonlinearity impacts the
inferred properties of individual voxels, and it has a plausible biological
interpretation. One benefit of quantitative encoding models is that estimated
models can be used to decode brain activity, in order to identify which
specific image was seen by an observer. Encoding models estimated by our
approach also improve such image identification by about 12\% when the correct
image is one of 11,500 possible images.
\end{abstract}

%

\end{frontmatter}
\setcounter{footnote}{7}

\section{Introduction}
\label{secintroduction}
One of the main differences between human brains and those of other
animals is the size of the neocortex [\citet{Frahm1982};
 \citet{Hofman1989};
  \citet{Rakic1995};
 \citet{VanEssen1997}]. Humans have one of the
largest neocortical sheets, relative to their body weight, in the
entire animal kingdom. The human neocortex is not a single
undifferentiated functional unit, but consists of several hundred
individual processing modules called areas. These areas are arranged in
a highly interconnected, hierarchically organized network. The visual
system alone consists of several dozen different visual \textit{areas},
each of which plays a distinct functional role in vision. The largest
visual area (indeed, the largest area in the entire neocortex) is the
primary visual cortex, area V1. Because of its central importance in
vision, area V1 has long been a primary target for computational modeling.

The most powerful tool available for measuring human brain activity is
 functional MRI (fMRI). However, fMRI data provide a rather
complicated window on neural function. First, fMRI does not measure
neuronal activity directly, but rather measures changes in blood
oxygenation caused by metabolic processes in neurons. Thus, fMRI
provides an indirect and nonlinear measure of neuronal activity.
Second, fMRI has a fairly low temporal and spatial resolution. The
temporal resolution is determined by physical changes in blood
oxygenation, which are two orders of magnitude slower than changes in
neural activity. The spatial resolution is determined by the physical
constraints of the fMRI scanner ({i.e}, limits on the strength of
the magnetic fields that can be produced, and limits on the power of
the radio frequency energy that can be deposited safely in the tissue).
In practice, fMRI signals usually have a temporal resolution of 1--2
seconds, and a spatial resolution of 2--4 millimeters. Thus, a typical
fMRI experiment might produce data from 30,000--60,000 individual
voxels ({i.e.}, volumetric pixels) every 1--2 seconds. These data
must first be filtered to remove nonstationary noise due to subject
movement and random changes in blood pressure. Then they can be modeled
and analyzed in order to address specific hypotheses of interest.

One recent approach for modeling fMRI data is to use a training data
set to estimate a separate model for each recorded voxel, and to test
predictions on a separate validation data set. In computational
neuroscience these models are called \textit{encoding} models, because
they describe how information about the sensory stimulus is encoded in
measured brain activity. Alternative hypotheses about visual function
can be tested by comparing prediction accuracy of multiple encoding
models that embody each hypothesis [\citet{Naselaris2010}].
Furthermore, estimated encoding models can be converted directly into
\textit{decoding} models, which can in turn be used to classify,
identify or reconstruct the visual stimulus from brain activity
measurements alone [\citet{Naselaris2010}]. These decoding models can
be used to measure how much information about specific stimulus
features can be extracted from brain activity measurements, and to
relate these measurement directly to behavior [\citet
{Raizada2010};
 \citet{Walther2009};
 \citet{Williams2007}].

Most encoding and decoding models rely on parametric regression methods that
assume the response is linearly related with stimulus features after fixed
parametric nonlinear transformation(s). These transformations may be
necessitated by nonlinearities in neural processes
[e.g., \citet{Carandini1997}], and other potential sources inherent to fMRI
such as dynamics of blood flow and oxygenation in the brain
[\citet{Buxton1998};
 \citet{Buxton2004}] and other biological factors
[\citet{Lauritzen2005}]. However, it can be difficult to guess the most
appropriate form of the transformation(s), especially when there are thousands
of voxels and thousands of features, and when there may be different
transformations for different features and different voxels. Inappropriate
transformations will most likely adversely affect prediction accuracy and
might also result in incorrect inferences and interpretations of the fitted
models.

\begin{figure}

\includegraphics{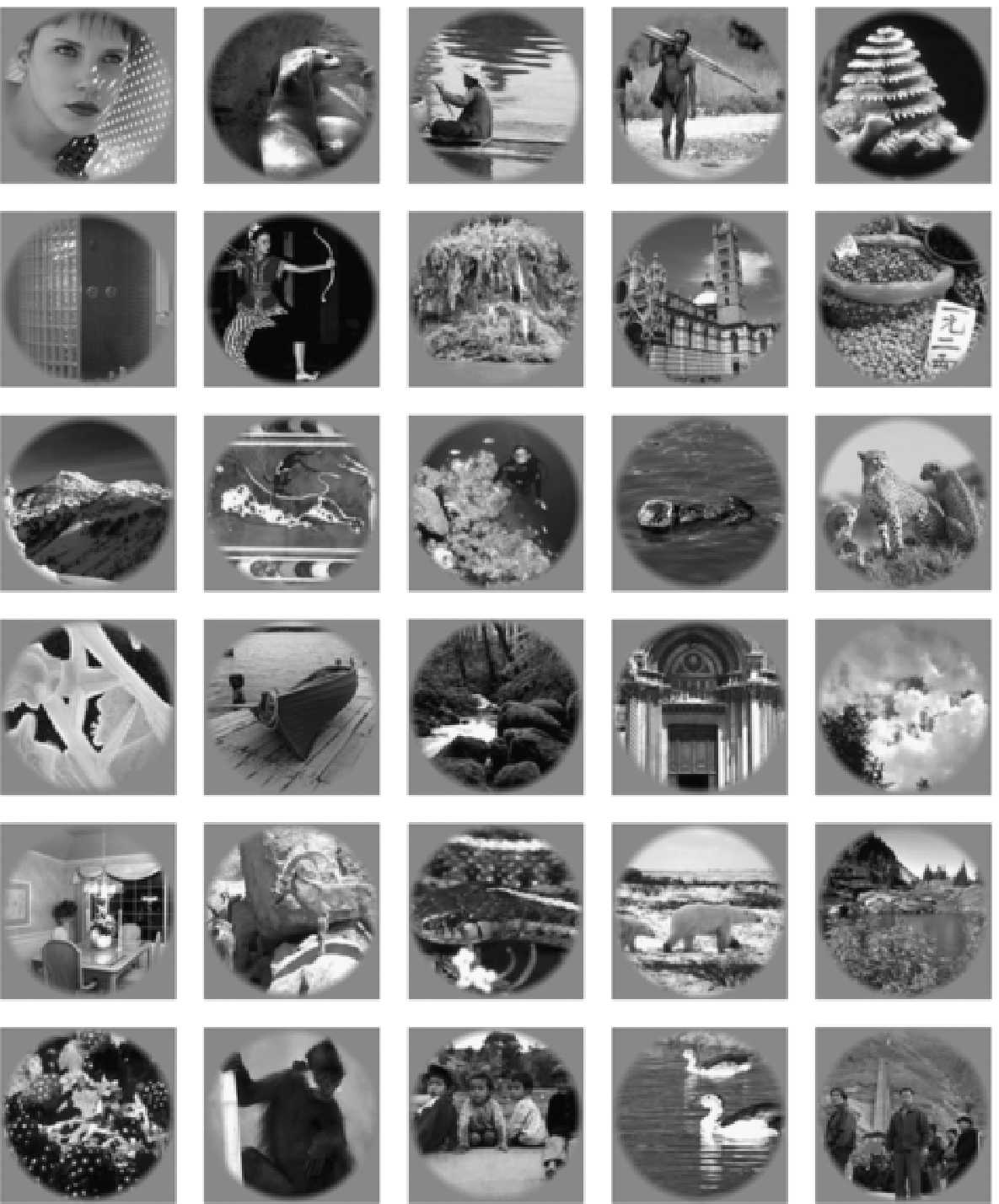}

\caption{Examples of natural image stimuli.
The natural images used in the experiment
were sampled from a large database of images obtained from a commercial
digital library (Corel Stock Photo Libraries from Corel Corporation). The
images covered $20 \times20$ degrees of the field of view, and were cropped
to a circular aperture and blended into the background to reduce edge
effects.}\label{fignatural-images}
\end{figure}

In this paper we use a new, sparse and flexible nonparametric approach
to more adequately model the nonlinearity in encoding models for fMRI
voxels in human area V1. The data were collected in an earlier study
[\citet{Kay2008}]. The stimuli were grayscale natural images (see
Figure~\ref{fignatural-images}). The original analysis focused on a
class of models that included a fixed parametric nonlinear
transformation of the stimuli, followed by linear weighting. Here we
show by residual analysis that this model does not account for a
substantial nonlinear response component (Section~\ref{secencoding}). We
therefore model these data by a sparse nonparametric method [\citet
{Ravikumar2009}] after preselection of features by marginal
correlation. The resulting model qualitatively affects inferred tuning
properties of V1 voxels (Section~\ref{sectuning}), and it substantially
improves response prediction (Section~\ref{secadditivemodels}). The
sparse nonparametric model also improves decoding accuracy (Section~\ref{secdecoding}). We conclude that the nonlinearities found in the
responses of voxels measured using fMRI impact both model performance
and model interpretation. Although our paper focuses entirely on area
V1, our approach can be extended easily to voxels recorded in other
areas of the brain.

\section{Background on V1}
\label{secbackground}
Brain area V1 is located in the occipital cortex and is an early
processing area of the visual pathway. It receives much of its input
from the lateral geniculate nucleus---a small cluster of cells in the
thalamus that is the brain's primary relay center for visual
information from the eye. Many of the properties of V1 neurons have
been described by visual neuroscientists [see \citet{Devalois1990} for
a summary]. In most cases these neurons are described as
spatio-temporal filters that respond whenever the stimulus matches the
\textit{tuning properties} of the filter. The important spatial tuning
properties for V1 neurons are related to spatial position, orientation
and spatial frequency. Thus, each V1 neuron responds maximally to
stimuli that appear at a particular spatial location within the visual
field, with a particular orientation and spatial frequency. Stimuli at
different spatial positions, orientations and frequencies will elicit
lower responses from the neuron. Because V1 neurons are tuned for
spatial position, orientation and spatial frequency they are often
modeled as Gabor filters (whose impulse response is the product of a
harmonic function and a Gaussian kernel) [\citet{Devalois1990}].

Although tuning for orientation and spatial frequency can be described
using a linear filter model, it is well established that individual V1
neurons do not behave exactly like linear filters. Studies using
white noise stimuli have reported a nonlinear relationship between
linear filter outputs and measured neural responses [e.g., \citet{Sharpee2008}; \citet
{Touryan2002}]. Furthermore, it is known that the responses
of V1 neurons saturate (like $\sqrt{x}$ or $\log{x}$) with increasing
contrast [e.g., \citet{Albrecht1982};
 \citet{Sclar1990}].
Finally, there is evidence that the responses of V1 neurons are
normalized by the activity of other neurons in their spatial or
functional neighborhood. This phenomenon---known as \textit{divisive
normalization}---can account for a variety of nonlinear behaviors
exhibited by V1 neurons [\citet{Carandini1997};
 \citet{Heeger1992}]. It is
reasonable to expect that the nonlinearities at the neural level will
affect voxel responses evoked by natural images, so a statistical model
should describe adequately these nonlinearities.

\section{The fMRI data}
\label{secthedata}

The data consist of fMRI measurements of blood oxygen level-dependent
activity (or BOLD response) at $m=1\mbox{,}331$ voxels in area V1 of a single
human subject [see \citet{Kay2008}]. The voxels, measuring $2 \times2
\times2.5$ millimeters, were acquired in coronal slices using a 4T
INOVA MR (Varian, Inc., Palo Alto, CA) scanner, at a rate of 1Hz, over
multiple sessions. Two sets of data were collected during the
experiment: training and validation. During the training stage the
subject viewed $n=1\mbox{,}750$ grayscale natural images randomly selected
from an image database, each presented twice (but not consecutively) in
a pseudorandom sequence; see Figure~\ref{fignatural-images}. Each image
was presented in an {ON-OFF-ON-OFF-ON} pattern for 1 second with an additional 3 seconds
OFF between presentations. For the validation data the subject viewed
$120$ novel natural images presented in the same way as in the training
stage, but with a total of $13$
presentations of each image. Data collection required approximately 10
hours in the scanner, distributed across 5 two hour sessions.

Data preprocessing is necessary to correct several sampling
artifacts\break
that~are intrinsic to fMRI. First, volumes were manually
co-registered\break
(in-house software) to correct for differences in head positioning
across\break sessions. Slice-timing and automated motion corrections (SPM99,\break
\url{http://www.fil.ion.ucl.ac.uk/spm}) were applied to volumes
acquired within the same session. These corrections are standard and
their details are explained in the supplementary information of \citet{Kay2008}.

Our encoding and decoding analyses depend upon defining a single scalar
fMRI voxel response to each image.
The procedures used to extract this scalar response from the BOLD time
series measurements acquired during the fMRI experiment are described
in the \hyperref[sechemodynamic]{Appendix}.
In short, we assume that each distinct image evokes a fixed timecourse
response, and that the response timecourses evoked by different images
differ by only a scale factor. We use a model in which the response
timecourses and scale factors are treated as separable parameters, and
then use these scale factors as the scalar voxel responses to each
image. By extracting a single scalar response from the entire
timecourse, we effectively separate the salient image-evoked attributes
of the BOLD measurements from those attributes due to the BOLD effect
itself [\citet{Kay2008hbm}].

\section{Encoding the V1 voxel response}
\label{secencoding}

An encoding model that predicts brain activity in response to stimuli is
important for neuroscientists who can use the model predictions to investigate
and test hypotheses about the transformation from stimulus to response.
In the
context of fMRI, the voxel response is a proxy for brain activity, and
so an
fMRI encoding model predicts voxel responses. Let $Y_v$ be the response of
voxel $v$ to an image stimulus $S$. We follow the approach of \citet
{Kay2008} and model the conditional mean
response,
\[
\mu_v(s) := \E(Y_v | S = s)
 ,
\]
as a function of local contrast energy features derived from projecting
the image onto a 2D Gabor wavelet basis. These features are inspired by
the known properties of neurons in V1, and are well established in
visual neuroscience [see, e.g., \citet
{Adelson1985};
 \citet{Jones1987};
 \citet{Olshausen1996}]. A 2D Gabor wavelet $g$ is the
pointwise product of a complex 2D Fourier basis function and a Gaussian kernel:
\[
g(a, b)
\propto
\exp (
2 \pi i \omega\tilde{a}
 )
\times
\exp \biggl(
-\frac{\tilde{a}^2}{2\sigma_1^2} - \frac{\tilde{b}^2}{2\sigma_2^2}
 \biggr),
 \]
 where
 \begin{eqnarray*}
\tilde{a} &=& (a - a_0) \cos\theta+ (a - a_0) \sin\theta,\\
\tilde{b} &=& (b - b_0) \cos\theta- (b - b_0) \sin\theta
 .
\end{eqnarray*}

%
\begin{figure}

\includegraphics{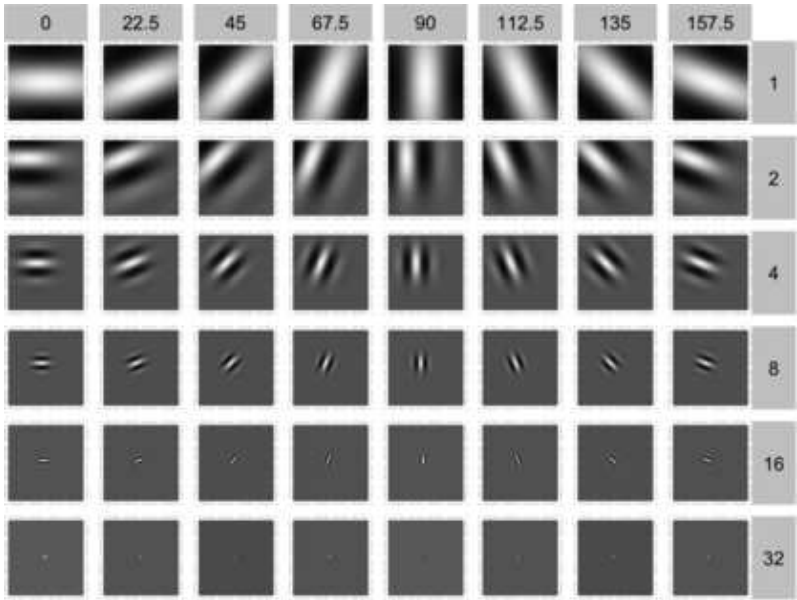}

\caption{Examples of Gabor wavelets.
The basis used by the encoding model is organized into 6
spatial scales (rows) and 8 orientations (columns). The imaginary part of
the wavelets is not shown.}
\label{figgabor-wavelets}
\end{figure}

The basis we use is organized into 6 spatial scales/frequencies
$(\omega, \sigma_1, \sigma_2)$, where wavelets tile spatial
locations $(a_0,b_0)$ and
8 possible orientations $\theta$, for a total of $
p
= (1^2 + 2^2 + 4^2 + 8^2 + 16^2 + 32^2) \times8
= 10\mbox{,}920
$ wavelets. Figure~\ref{figgabor-wavelets} shows all of the possible scale
and orientation pairs.

Let $g_j$ denote a wavelet in the basis. The local contrast energy feature
is defined as
\[
X_j(s)
:=  \biggl[ \sum_{a,b} \Re g_j(a,b) s(a,b)  \biggr]^2
+  \biggl[ \sum_{a,b} \Im g_j(a,b) s(a,b)  \biggr]^2
\]
for $j = 1,\ldots, p=10\mbox{,}920$.
The feature set is essentially a localized version of the (estimated)
Fourier power spectrum of the image. Each feature measures the amount
of contrast energy in the image at a particular frequency, orientation
and location.

\subsection{Sparse linear models}
\label{seclinearmodels}
The model proposed in \citet{Kay2008} assumes that $\mu_v(s)$ is a weighted
sum of a fixed transformation of the local contrast energy features. They
applied a square root transformation to $X_j$ to make the relationship between
$\mu_v(s)$ and the transformed features more linear. Thus, their model is\looseness=1
%
\begin{equation}
\label{eqsqrt-model}
\mu_v(s) = \beta_{v0} + \sum_{j=1}^p \beta_{vj} \sqrt{X_j(s)}
 .
\end{equation}
We refer to \eqref{eqsqrt-model} as the $\mathit{sqrt}(X)$ model.
\citet{Kay2008} fit this model separately for each of the $1\mbox{,}331$ voxels,
using gradient descent on the squared error loss with early stopping
[see, e.g., \citet{Friedman2004}], and demonstrated that the fitted
models could be used to identify, from a large set of novel images, which
specific image had been viewed by the subject. They used a simple decoding
method that selects, from a set of candidates, the image $s$ whose predicted
voxel response pattern $(\hat{\mu}_v(s) \dvtx  v = 1,2,\ldots)$ is most correlated
with the observed voxel response pattern $(Y_v \dvtx  v = 1,2,\ldots)$.
Although \citet{Kay2008} focused on decoding, the encoding model is
clearly an integral part of their approach. We found a substantial nonlinear
aspect of the voxel response that their encoding $\mathit{sqrt}(X)$ model does not
take into
account.

\begin{figure}

\includegraphics{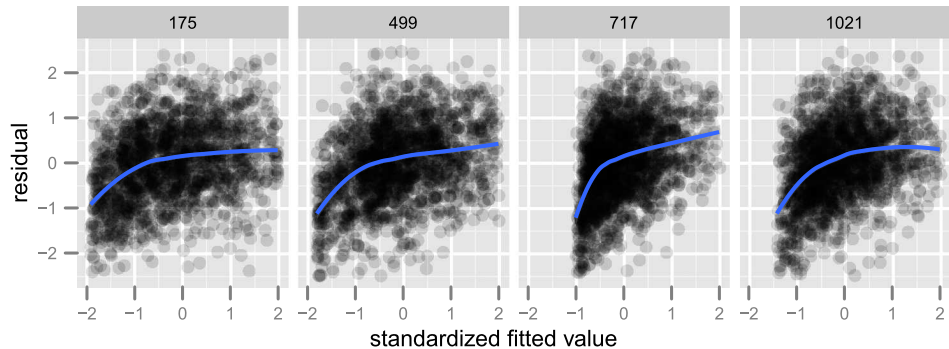}

\caption{Residual and fitted values of model \protect\eqref{eqsqrt-model} for
four different voxels (labeled above). The solid curves show a LOESS fit
of the residual on the fitted values.}
\label{figlinear-voxel-residuals}
\end{figure}

\begin{figure}

\includegraphics{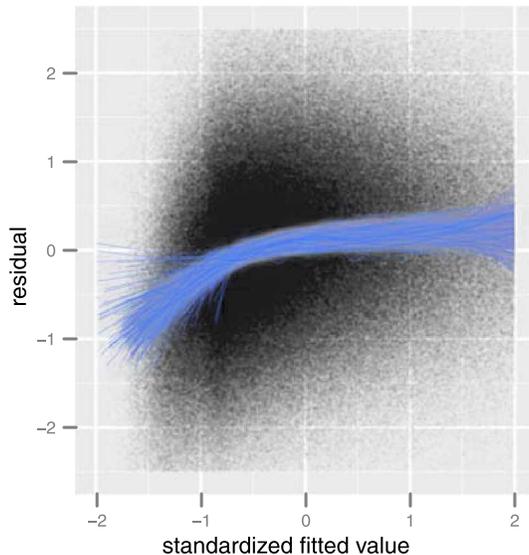}

\caption{Residual and standardized fitted values of model \protect\eqref{eqsqrt-model}
blended across all $1\mbox{,}331$ voxels. The solid curves show the LOESS fits of
the residuals on the fitted values for each voxel.}
\label{figlinear-all-residuals}
\end{figure}

Since the gradient descent method with early stopping is closely
related to the Lasso method [\citet{Friedman2004}],
we fit the model \eqref{eqsqrt-model} separately to each voxel [as in
\citet{Kay2008}] using Lasso [\citet{Tibshirani1996}], and selected the
regularization parameters with BIC (using the number of nonzero coefficients
in a Lasso model as the degrees of freedom).
Figure~\ref{figlinear-voxel-residuals} shows plots of the residuals and fitted
values for four different voxels. With the aid of a LOESS smoother
[\citet{Cleveland1988}], we see a nonlinear relationship between the residual
and the fitted values. This pattern is not unique to these four voxels. We
extended this analysis to all $1\mbox{,}331$ voxels. By standardizing the fitted
values, we can overlay the smoothers for all $1\mbox{,}331$ voxels and inspect for
systematic deviations from the $\mathit{sqrt}(X)$ model across all voxels.
Figure~\ref{figlinear-all-residuals} shows the\vadjust{\goodbreak} result. Nonlinearity
beyond the $\mathit{sqrt}(X)$ model is present
in almost all voxels, and, moreover, the residuals appear to be
heteroskedastic.

%

Composing the square root transformation with an additional nonlinear
transformation could absorb some of the residual nonlinearity in the
$\mathit{sqrt}(X)$ model.
Instead of the square root, $\log(1+\sqrt{x})$ was used
by \citet{Naselaris2009} to analyze the same data set as we do in this paper
and it has also been used in other applications [see \citet
{Kafadar2006} for an example in the analysis of internet traffic data].
The resulting model is
%
\begin{equation}
\label{eqlog-model}
\mu_v(s)
= \beta_{v0} + \sum_{j=1}^p \beta_{vj} \log \bigl(1 + \sqrt
{X_j(s)} \bigr)
 ,
\end{equation}
and we refer to it as the $\mathit{log}(1+\mathit{sqrt}(X))$ model.

\begin{figure}

\includegraphics{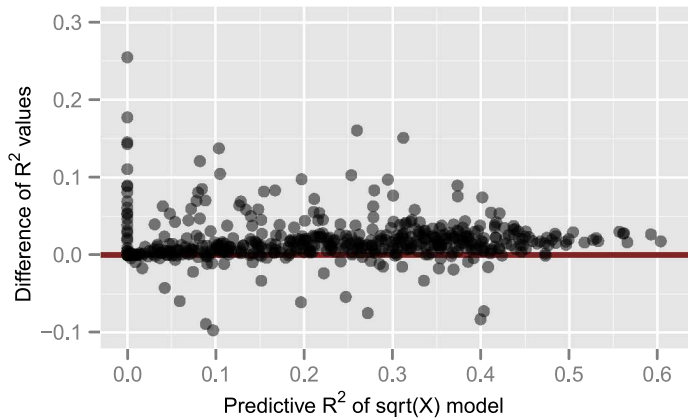}

\caption{Comparison of voxel-wise predictive $R^2$ (based on the validation data)
of
the $\mathit{log}(1+ \mathit{sqrt}(X))$ model \protect\eqref{eqlog-model}
and the $\mathit{sqrt}(X)$ model \protect\eqref{eqsqrt-model}.
 The vertical
axis shows the difference $R^2$ of \protect\eqref{eqlog-model} $- R^2$ of
\protect\eqref{eqsqrt-model}. The median improvement of model
\protect\eqref{eqlog-model} is $5.5\%$ for voxels where both models have a
predictive $R^2 > 0.1$.}
\label{figcc-sqrtx-logx}
\end{figure}

We fit model \eqref{eqlog-model} using Lasso with BIC, and compared its
prediction performance with model \eqref{eqsqrt-model} by evaluating the
squared correlation (predictive $R^2$) between the predicted and actual
response across all 120 images in the validation set.
Figure~\ref{figcc-sqrtx-logx} shows the difference in predictive $R^2$
values of
the two models for each voxel. There is an improvement in prediction
performance (median $5.5\%$ for voxels where both models have an $R^2 > 0.1$)
with model \eqref{eqlog-model}. However, examination of residual plots
(not shown) reveals that there is still residual nonlinearity.

\subsection{Sparse additive (nonparametric) models}
\label{secadditivemodels}
The $\sqrt{x}$ and $\operatorname{log}(1+\sqrt{x})$ transformations were used in previous
work to approximate the contrast saturation of the BOLD response.
Rather than
trying other fixed transformations to account for the
nonlinearities in the voxel response, we employed a sparse nonparametric
approach that is based on the additive model. The additive model
[cf. \citet{Hastie1999}] is a useful generalization of the linear
model that allows the feature transformations to be estimated from the data.
Rather than assuming that the conditional mean $\mu$ is a linear
function (of
fixed transformations) of the features, the additive (nonparametric)
model assumes that
%
\begin{equation}
\mu= \beta_{0} + \sum_{j=1}^p f_{j}(X_j)
 ,
\label{eqgam}
\end{equation}
where $f_{j} \in\mathcal{H}_j$ are unknown, mean 0 predictor
functions in
some Hilbert spaces $\mathcal{H}_j$. The linear model is a special
case where
the predictor functions are assumed to be of the form $f_{j}(x) = \beta_{j}
x$. The monograph of Hastie and Tibshirani describes methods of estimation
and algorithms for fitting \eqref{eqgam}, however, the setting there
is more
classical in that the methods are most appropriate for low-dimensional
problems (small $p$, large $n$).

\citet{Ravikumar2009} extended the additive model methodology to the
high-dimensional setting by incorporating ideas from the Lasso.
Their sparse additive model (SPAM) adds a sparsity assumption to \eqref{eqgam} by assuming that the set of active predictors $\{j\dvtx  f_j \neq
0\}$ is sparse. They propose fitting \eqref{eqgam} under this sparsity
assumption by minimization of the penalized squared error loss
%
\begin{equation}
\min_{f_j \in\hat{\mathcal{H}}_j, \beta_0}
\Biggl\|\mathbf{Y} - \beta_0\mathbf{1} - \sum_{j=1}^p f_{j}(\mathbf
{X}_j) \Biggr\|^2
+ \lambda\sum_{j=1}^p \|f_{j}(\mathbf{X}_j)\|
 ,
\label{eqspamloss}
\end{equation}
where $\|\cdot\|$ is the Euclidean norm in $\Real^n$,
$\mathbf{Y}$ is the $n$-vector of sample responses, $\mathbf{1}$~is the
vector of $1$'s, $f_j(\mathbf{X}_j)$ is the vector obtained by
applying $f_j$
to each sample of $X_j$, and $\lambda\geq0$. The penalty term,
$\lambda
\sum_{j=1}^p \|f_j(\mathbf{X}_j)\|$, is the functional equivalent
of the Lasso penalty. It simultaneously encourages sparsity (setting many
$f_j$ to zero) and shrinkage of the estimated predictor functions
by acting as an L1 penalty on the empirical L2 function norms $\|
f_j(\mathbf{X}_j)\|$, $j=1,\ldots,p$.
%
\begin{figure}[b]
\hrule
\vskip0.25em
{\fontsize{9pt}{11pt}\selectfont{\begin{algorithmic}
\REQUIRE Sample vectors $(\mathbf{Y}$, $\mathbf{X}_1,\ldots, \mathbf{X}_p)$,
smoothers ($\smooth_1, \ldots, \smooth_p$), and regularization parameter
($\lambda\geq0$)
\STATE$\hat{\beta}_0 \leftarrow\bar{\mathbf{Y}}$\vspace*{1pt}
\STATE$\hat{f}_j \leftarrow0$ for $j = 1,\ldots, p$
\REPEAT
\FOR{$j = 1$ to $p$}
\STATE$\mathbf{R}_j \leftarrow\mathbf{Y} - \hat{\beta}_0\mathbf
{1} - \sum_{k \neq j} \hat{f}_k(\mathbf{X_k})$---compute the partial residual
\STATE$s_j \leftarrow\smooth_j(\mathbf{R}_j)$\vspace*{1pt}
\STATE$\hat{f}_j \leftarrow s_j (1 - \lambda/\|s_j\|)_+$---soft-threshold
\ENDFOR
\UNTIL{$RSS = \|\mathbf{Y} - \hat{\beta}_0\mathbf{1} - \sum_j
\hat{f}_j(\mathbf{X}_j)\|^2$ converges
}
\RETURN estimated intercept $\hat{\beta}_0$ and predictor functions
$\hat{f}_1, \hat{f}_2, \ldots, \hat{f}_p$
\end{algorithmic}}}
\vskip0.25em
\hrule
\caption{The SPAM backfitting algorithm.}
\label{figspam-backfitting}
\end{figure}
The algorithm proposed by \citet{Ravikumar2009} for solving the sample
version of the SPAM
optimization problem \eqref{eqspamloss} is shown in
Figure~\ref{figspam-backfitting}. It generalizes the well-known backfitting
algorithm [\citet{Friedman1981}] by incorporating an additional
soft-thresholding step. The main bottleneck of the algorithm is the complexity
of the smoothing step.

We did not apply SPAM directly to the feature $X_j(s)$, but instead
applied it to the transformed feature, $\log(1 + \sqrt{X_j(s)})$. We
refer to the model
%
\begin{equation}
\mu_v(s)
= \beta_{v0}
+ \sum_{j=1}^p f_{vj} \bigl(\log \bigl(1 + \sqrt{X_j(s)} \bigr) \bigr)
\label{eqvspam}
\end{equation}
as V-SPAM---``V'' for visual cortex and V1 neuron-inspired features.
There is
no loss in generality of this model when compared with \eqref{eqgam}, but
there is a practical benefit because the $\log(1 + \sqrt{X_j(s)})$ feature
tends to be better spread out than the $X_j(s)$ feature. This has a direct
effect on the smoothness of $f_{vj}$. Although\vspace*{1pt} we did not try other
transformations, we found that applying the SPAM model directly to the
$X_j(s)$ features rather than $\log(1+\sqrt{X_j(s)})$ resulted in poorer
fitting models.

We fit the V-SPAM model separately to each voxel, using cubic spline
smoothers for the $f_{vj}$. We placed knots at the deciles of the $\log
(1+\sqrt{X_j})$ feature distributions and fixed the effective degrees
of freedom [trace of the corresponding smoothing
matrix; {cf.} \citet{Hastie1999}] to 4 for each smoother. This choice was based on
examination of a few partial residual plots from model \eqref{eqlog-model}
and comparison of smooths for different effective degrees of freedoms.
We felt
that optimizing the smoothing parameters across features and voxels (with
generalized cross-validation or some other criterion) would add too much
complexity and computational burden to the fitting procedure.

The amount of time required to fit the V-SPAM model for a single voxel
with $10\mbox{,}920$ features is considerably longer than for fitting a linear
model, because of the complexity of the smoothing step. So for
computational reasons we reduced the number of features to 500 by
screening out those that have low marginal correlation with the
response, which reduced the time to fit one voxel to about 10
seconds.\footnote{Timing for an 8-core, 2.8 GHz Intel Xeon-based
computer using a multithreaded linear algebra library with software
written in R.} We selected the regularization parameter $\lambda$
using BIC with the degrees of freedom of a candidate model defined to
be the sum of the effective degrees of freedom of the active smoothers
(those corresponding to nonzero estimates of $f_j$).

\begin{figure}

\includegraphics{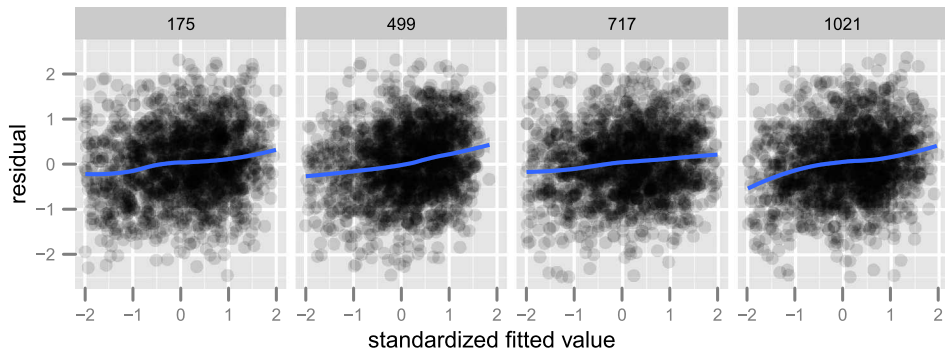}

\caption{Residual and fitted values of V-SPAM \protect\eqref{eqvspam} for four
different voxels (labeled above). The solid curves show a LOESS fit of the
residual on the fitted values. Compare with
Figure~\protect\ref{figlinear-voxel-residuals}. The linear trend in the
residuals is
due to the shrinkage effect of the penalty in the SPAM criterion
\protect\eqref{eqspamloss}.}
\label{figvspam-voxel-residuals}
\end{figure}

\begin{figure}

\includegraphics{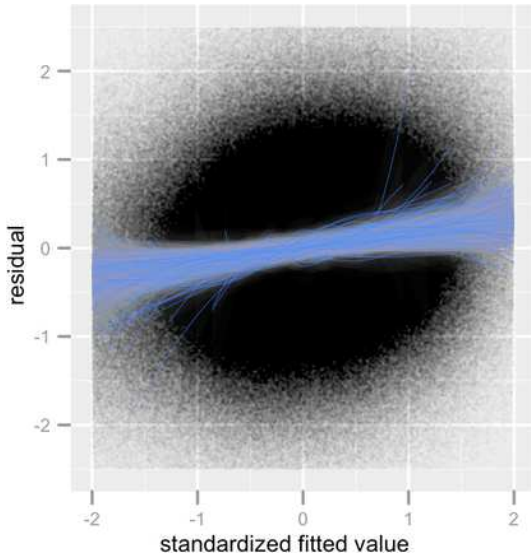}

\caption{Residual and standardized fitted values of V-SPAM \protect\eqref{eqvspam} for all
$1\mbox{,}331$ voxels. The solid curves show the LOESS fits of the residuals on
the fitted values for each voxel. Compare with
Figure~\protect\ref{figlinear-all-residuals}.}
\label{figvspam-all-residuals}
\end{figure}

Figure~\ref{figvspam-voxel-residuals} shows residual and fitted value plots
for the four voxels that we examined in the previous section. Little
residual nonlinearity remains in this aspect of the V-SPAM fit. The residual
linear trend in the LOESS curve is due to the shrinkage effect of the SPAM
penalty---the residuals of a penalized least squares fit are necessarily
correlated with the fitted values. Figure~\ref{figvspam-all-residuals} shows
the residuals and fitted values of V-SPAM for all $1\mbox{,}331$ voxels. In contrast
to Figure~\ref{figlinear-all-residuals}, there is neither a visible pattern
of nonlinearity, nor a visible pattern of heteroskedasticity.

\begin{figure}
\centering
\begin{tabular}{c}

\includegraphics{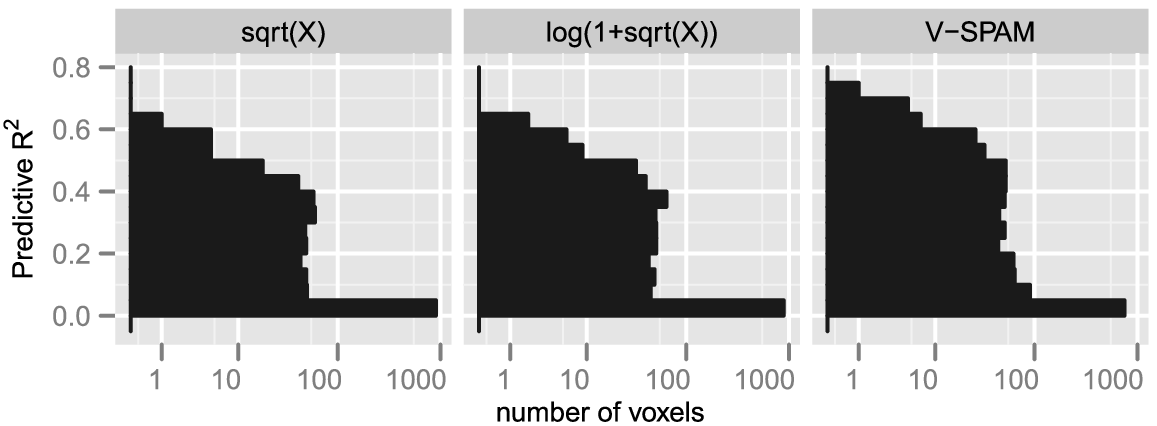}
\\
(a)\\[6pt]
{
\includegraphics{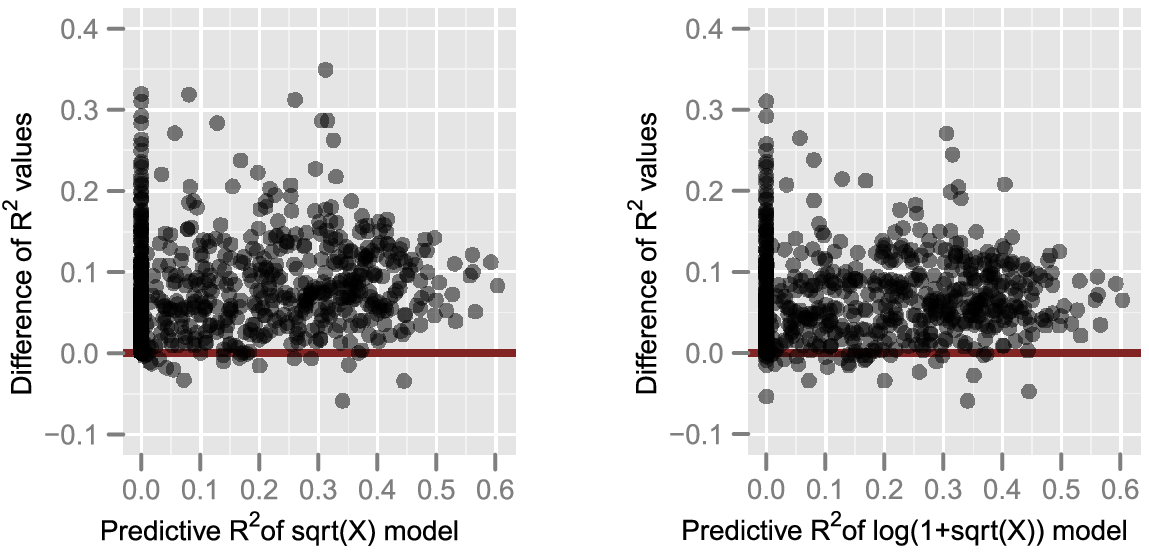}
}\\
(b)
\end{tabular}
\caption{Comparison of voxel-wise predictive $R^2$ (based on the validation data)
of
the $\mathit{sqrt}(X)$ model \protect\eqref{eqsqrt-model}, the $\mathit{log}(1+\mathit{sqrt}(X))$
model \protect\eqref{eqlog-model}
 and V-SPAM \protect\eqref{eqvspam}.
\textup{(a)} Histograms of the predictive $R^2$ value across voxels. They
are displayed sideways to ease comparison.
\textup{(b)} Difference of predictive $R^2$ values of V-SPAM \protect\eqref{eqvspam}:
(left) $\mathit{sqrt}(X)$ model \protect\eqref{eqsqrt-model}; (right) $\mathit{log}(1+\mathit{sqrt}(X))$
model \protect\eqref{eqlog-model}.}
\label{figcc-vspam}
\end{figure}

The V-SPAM model better addresses nonlinearities in the voxel response. To
determine if this model leads to improved prediction performance, we examined
the squared correlation (predictive $R^2$) between the predicted and actual
response across all 120 images in the validation set. Figure~\ref{figcc-vspam} compares the predictive $R^2$ of the V-SPAM model for
each voxel with those of the $\mathit{sqrt}(X)$ model \eqref{eqsqrt-model} and the
$\mathit{log}(1+\mathit{sqrt}(X))$ model \eqref{eqlog-model}. Across most voxels, there
is a
substantial improvement in prediction performance. The median (across voxels
where both models have a predictive $R^2 > 0.1$) is $26.4\%$ over the $\mathit{sqrt}(X)$
model, and $19.9\%$ over the $\mathit{log}(1+\mathit{sqrt}(X))$ model. Thus, the
additional nonlinear
aspects of the response revealed in the residual plots (Figures \ref{figlinear-voxel-residuals} and \ref{figlinear-all-residuals})
for the parametric $\mathit{sqrt}(X)$ and $\mathit{log}(1+\mathit{sqrt}(X))$ models are real and they
account for a substantial part of the
prediction of the voxel response.

\section{Decoding the V1 voxel response}
\label{secdecoding}
Decoding models have received a great deal of attention recently
because of their role in potential ``mind reading'' devices. Decoding
models are also useful from a statistical point of view because their
results can be judged directly in the known and controlled stimulus space.
Here we show that accurately characterizing nonlinearities with the
V-SPAM encoding model (presented in the preceding section) leads to
substantially improved decoding.

We used a Naive Bayes approach similar to that proposed by \citet
{Naselaris2009} to derive a decoding model from the V-SPAM encoding
model. Recall that $Y_v$ ($v = 1, \ldots, m$ and $m = 1\mbox{,}331$) is the
response of voxel $v$ to image $S$. A simple model for $Y_v$ that is
compatible with the least squares fitting in Section~\ref{secencoding}
assumes that the conditional distribution of $Y_v$ given $S$ is Normal
with mean $\mu_v(S)$ and variance $\sigma_v^2$,
and that $Y_1, \ldots, Y_m$ are conditionally independent given $S$.
To complete the specification of the joint distribution of the stimulus
and response, we take an empirical approach [\citet{Naselaris2009}] by
considering a large collection of images $\bagofimages$ similar to
those used to acquire training and validation data.
The bag of images prior places equal probability on each image in
$\bagofimages$:
\[
\P(S = s) =
\cases{\displaystyle
\frac{1}{|\bagofimages|} ,&\quad  if $s \in\bagofimages$,\vspace*{3pt}\cr\displaystyle
0 ,&\quad  otherwise.
}
\]
This distribution only implicitly specifies the statistical structure of
natural images. With Bayes' rule we arrive at the decoding model
\[
p(s | y_1,\ldots, y_s)
\propto\exp \Biggl\{
- \sum_{v = 1}^m \frac{( y_v - \mu_v(s) )^2}{2\sigma_v^2}
 \Biggr\} \times\P(S = s)
 .
\]
This model suggests that we can identify the image $s$ that most
closely matches a given voxel response pattern $(Y_1, \ldots, Y_m)$ by
the rule
%
\begin{equation}
\argmax_s p(s | y_1,\ldots, y_s)
=
\argmin_{s \in\bagofimages} \sum_{v = 1}^m \frac{1}{\sigma_v^2} \bigl(
y_v - \mu_v(s) \bigr)^2
 .
\label{eqbayesrule}
\end{equation}
The fitted models from Section~\ref{secencoding} provide estimates of
$\mu_v$.
Given $\hat{\mu}_v$, the variance $\sigma_v^2$ can be estimated by
\[
\hat{\sigma}_v^2
= \frac{\|\mathbf{Y}_v - \hat{\mu}_v(\mathbf{S})\|^2}{n - \df
(\hat{\mu}_v)}
 ,
\]
where $\df(\hat{\mu}_v)$ is the degrees of freedom of the estimate
$\hat{\mu}_v$ (the number of nonzero coefficients in the case of
linear models, or 4 times the number of nonzero functions in the case
of V-SPAM; {cf.} Section~\ref{secadditivemodels}). Substituting
these estimates into
\eqref{eqbayesrule} gives the decoding rule
\[
\argmin_{s \in\bagofimages}
\sum_{v = 1}^m
\frac{1}{\hat{\sigma}_v^2} \bigl( y_v - \hat{\mu}_v(s) \bigr)^2
 .
\]
Although we have estimates for
every voxel, not every voxel may be useful for decoding---$\hat{\mu}_v$
may be a poor estimate of $\mu_v$ or $\mu_v(s)$ may be close to
constant for
every $s$. In that case, we may want to select a subset of voxels
$\usefulvoxels\subseteq\{1,\ldots,m\}$ and restrict the summation in the
above display to $\usefulvoxels$. Thus, we propose the decoding rule
%
\begin{equation}
\label{eqdecoding-model}
\hat{S}_\usefulvoxels(y_1, \ldots, y_m | \bagofimages)
=
\argmin_{s \in\bagofimages}
\sum_{v \in\usefulvoxels}
\frac{1}{\hat{\sigma}_v^2}\bigl ( y_v - \hat{\mu}_v(s) \bigr)^2
 .
\end{equation}
One strategy for voxel selection is to set a threshold $\alpha$ for
entry to $\usefulvoxels$ based on the usual $R^2$ computed with the
training data,
%
\begin{equation}
\label{eqtraining-R2}
\mbox{training } R^2(v)
=
1 -
\frac{\|\mathbf{Y}_v - \hat{\mu}_v(\mathbf{S})\|^2}
{\|\mathbf{Y}_v - \bar{\mathbf{Y}}_v\|^2}
 ,
\end{equation}
so that $\usefulvoxels_\alpha= \{v \dvtx  \mbox{training } R^2(v) >
\alpha\}$.
We will examine this strategy later in the section.

To use \eqref{eqdecoding-model} as a general purpose decoder, the
collection of images $\bagofimages$
should ideally be large enough so that every natural image $S$ is
``well-approximated'' by some image in $\bagofimages$. This requires a
distance function over natural images in order to formalize
``well-approximate,''
but it is not clear what the distance function should be. We consider
instead the
following paradigm. Suppose that the image stimulus $S$ that evoked the voxel
response pattern is actually contained in $\bagofimages$. Then it may be
possible for \eqref{eqdecoding-model} to recover $S$ exactly. This is
the basic premise of the identification problem where we ask if the
decoding rule can correctly identify $S$ from a set of candidates
$\bagofimages\cup\{S\}$. Within this paradigm, we assess \eqref{eqdecoding-model} by its \textit{identification error rate},
%
\begin{equation}
\label{eqid-error-rate}
\mbox{id error rate}
:=
\P \bigl(
\hat{S}_\usefulvoxels(Y_1', \ldots, Y_m' |
\bagofimages\cup\{S'\} ) \neq S'
  |
\hat{S}_\usefulvoxels(\cdots)
 \bigr)
 ,
\end{equation}
on a future stimulus and voxel response pair $\{S', (Y_1', \ldots,
Y_m')\}$ that is independent of the training data.

The identification error rate should increase as $|\bagofimages| = b$
increases. However, the rate at which it increases will depend on the model
used for estimating $\hat{\mu}_v$. We investigated this by starting with
a database $\imagedatabase$ of $11\mbox{,}499$ images (as in Figure~\ref{fignatural-images}) that are similar to, but do not include, the
images in the training data or validation data, and then repeating the
following experiment for different choices of $b$:
\begin{longlist}[(3)]
\item[(1)] Form $\bagofimages$ by drawing a sample of size $b$ without
replacement from $\imagedatabase$.
\item[(2)] Estimate the identification error rate \eqref{eqid-error-rate} using
the 120 stimulus and voxel response pairs $\{S', (Y_1',\ldots,Y_m') \}
$ in
the validation data.
\item[(3)] Average the estimated identification error rate over all possible
$\bagofimages\subseteq\imagedatabase$ of size $b$.
\end{longlist}
The average identification error rate can be computed without resorting
to Monte Carlo.
Given $\{S', (Y_1', \ldots, Y_m')\}$,
%
\begin{equation}
\label{eqcorrect-event}
\hat{S}_\usefulvoxels(Y_1', \ldots, Y_m' | \bagofimages\cup\{S'\}
) = S'
\end{equation}
if and only if
%
\begin{equation}
\label{eqscore}
\sum_{v\in\usefulvoxels}
\frac{1}{\hat{\sigma}_v^2} \bigl( Y_v' - \hat{\mu}_v(S) \bigr)^2
<
\sum_{v\in\usefulvoxels}
\frac{1}{\hat{\sigma}_v^2} \bigl( Y_v' - \hat{\mu}_v(s) \bigr)^2
\end{equation}
for every $s \in\bagofimages$. Since $\bagofimages$ is drawn by a simple
random sample, the number of times that event \eqref{eqscore} occurs
follows a hypergeometric distribution. So the conditional probability that
\eqref{eqcorrect-event} occurs is just the probability that a hypergeometric
random variable is equal to $b$. The parameters of this
hypergeometric distribution are given by the number of images in
$\imagedatabase$ that satisfy \eqref{eqscore}, the number of images in
$\imagedatabase$ that do not satisfy \eqref{eqscore}, and $b$.
Counting the
number of images in $\imagedatabase$ that satisfy/do not satisfy
\eqref{eqscore} is easy and only has to be done once for each $S$ in
the validation data,
regardless of $b$. Thus, the computation involves evaluating
\eqref{eqscore} $120 \times11\mbox{,}499$ times (since there are $120$
images in the validation data and $11\mbox{,}499$ images in $\imagedatabase
$), and then evaluating $120$ hypergeometric probabilities for each $b$.

\begin{figure}

\includegraphics{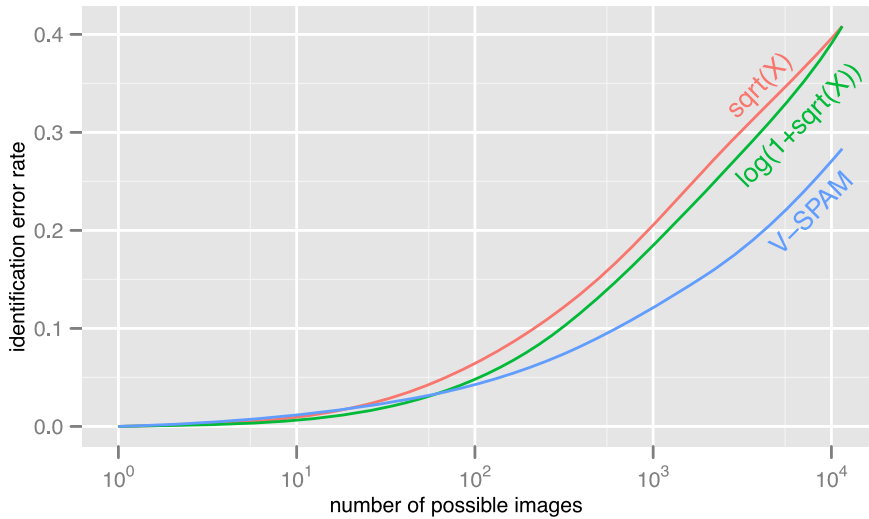}

\caption{Estimated average identification error rate \protect\eqref{eqid-error-rate} as a
function of the number of possible images ($|\bagofimages|+1$). The error
rates were estimated using the validation data and $\mathcal{B}$ randomly
sampled from a database of $11\mbox{,}499$ images.}
\label{figdecoding-error-numberofimages}
\end{figure}

Figure~\ref{figdecoding-error-numberofimages} shows the results of
applying the
preceding analysis to the fixed transformation models \eqref{eqsqrt-model}
and \eqref{eqlog-model} and the V-SPAM model \eqref{eqvspam}. Each
model has
its own subset of voxels $\usefulvoxels$ used by the decoding rule. We set
the training $R^2$ thresholds \eqref{eqtraining-R2} so that the corresponding
decoding rule used $|\usefulvoxels| = 400$ voxels for each model. When
$|\bagofimages|$ is small, identification is easy and all three models
have very low error rates. As the number of possible images increases, the
error rates of all three models increase but at different rates. At maximum,
when $\mathcal{B} = \mathcal{D}$ and there are $11\mbox{,}499 + 1 = 11\mbox{,}500$ candidate
images ($11\mbox{,}499$ images in $\mathcal{D}$ plus $1$ correct image not in
$\mathcal{D}$) for the decoding rule to choose from, the fixed
transformation models have an error rate of
about $40\%$, while the V-SPAM model has an error rate of about $28\%$.

\begin{figure}
\centering
\begin{tabular}{c}

\includegraphics{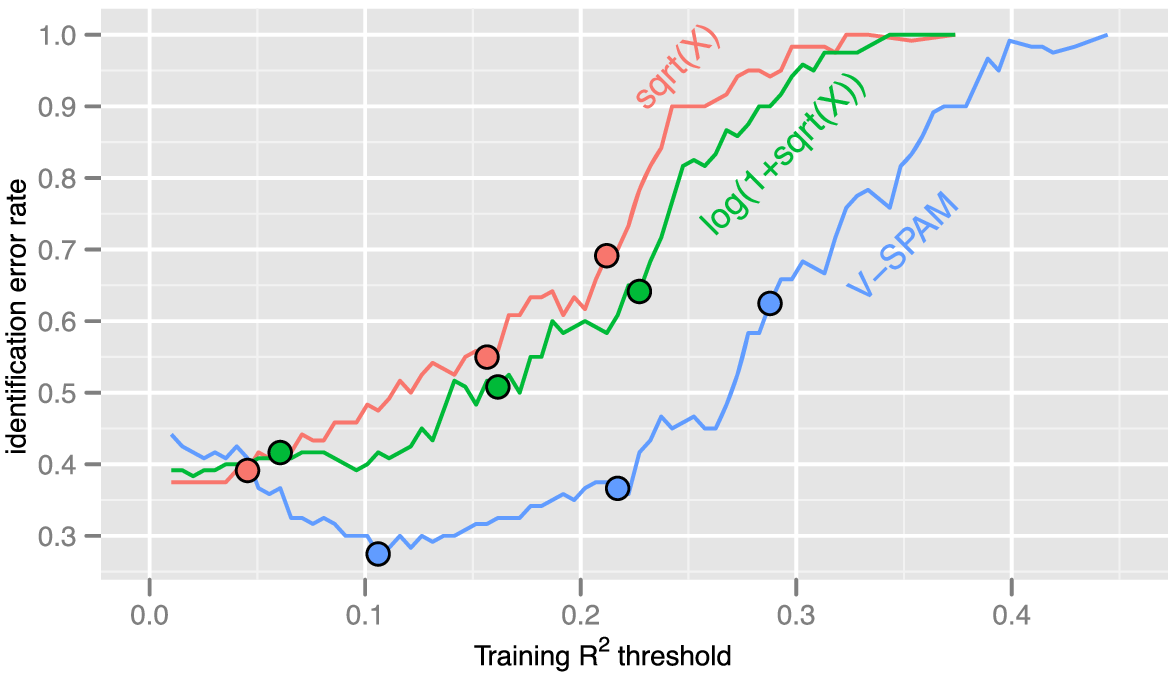}
\\
(a)\\[6pt]

\includegraphics{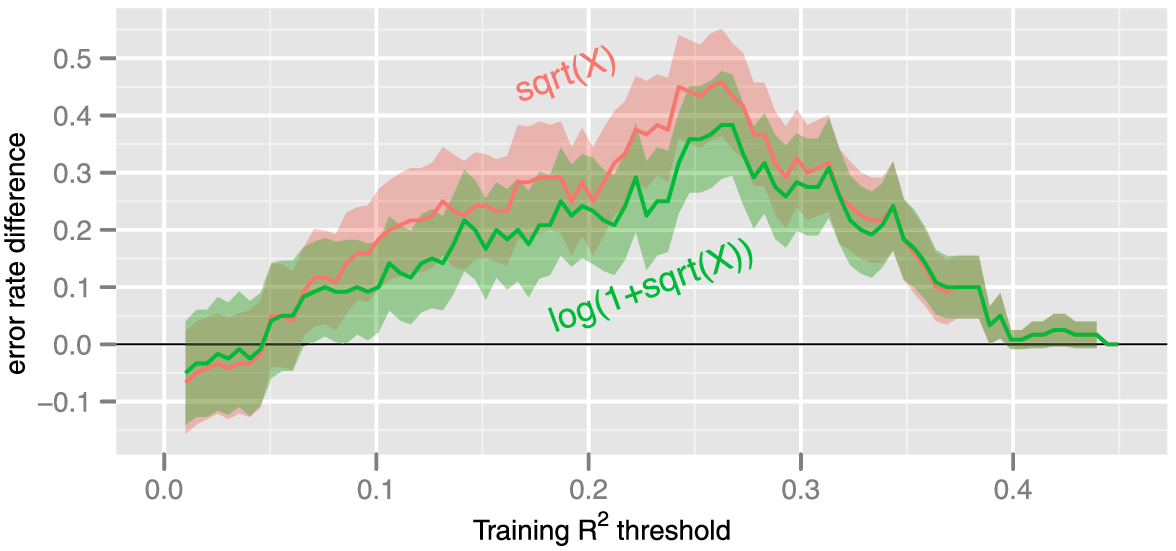}
\\
(b)
\end{tabular}
\caption{Identification error rate \protect\eqref{eqid-error-rate} as a function of the
training $R^2$ threshold \protect\eqref{eqtraining-R2} when the number of
possible images is $11\mbox{,}499+1$.
\textup{(a)}
Estimated identification error rate. The solid
circles on each curve mark the points where the number of voxels
used by the decoding rule is (from left to right) $400$, $200$ or $100$.
\textup{(b)}
Pointwise 95\% confidence bands for the difference between the
identification error rates of (upper) $\mathit{sqrt}(X)$ model \protect\eqref{eqsqrt-model}
and V-SPAM; (lower) $\mathit{log}(1+\mathit{sqrt}(X))$ model \protect\eqref{eqlog-model} and
V-SPAM. The confidence bands reflect uncertainty due to sampling variation
of the validation data.}
\label{figdecoding-error-r2}
\end{figure}
The ordering of and large gap between the fixed transformation models
and V-SPAM at maximum
does not depend on our choice of $|\mathcal{V}| = 400$ voxels. Fixing
$\mathcal{B} = \mathcal{D}$ so that the number of possible images is maximal,
we examined how the identification error rate varies as the training $R^2$
threshold is varied. Figure~\ref{figdecoding-error-r2} shows our results.
The threshold corresponding to $400$ voxels is larger for V-SPAM than
the fixed transformation models. It is about $0.1$ for V-SPAM and
$0.05$ for the fixed transformation models.
When the threshold is below $ 0.05$, the error rates of the three models are
indistinguishable. Above $ 0.05$, V-SPAM generally has a much lower error rate
than the fixed transformation models. In panel (a) of Figure~\ref{figdecoding-error-r2} we also see that
V-SPAM can achieve an error rate lower than the best of the fixed
transformation models with half as many voxels ($\leq200$ versus $\geq400$).
These results show that the substantial improvements in voxel
response prediction by V-SPAM can lead to substantial improvements in decoding
accuracy.

\section{Nonlinearity and inferred tuning properties}
\label{sectuning}
In computational neuroscience, the \textit{tuning function} describes
how the output of a neuron or voxel varies as a function of some
specific stimulus feature [\citet{Zhang1999}]. As such, the tuning
function is a special case of an encoding model, and once an encoding
model has been estimated, a tuning function can be extracted from the
model by integrating out all of the stimulus features except for those
of interest. In practice, this extraction is achieved by using an
encoding model to predict responses to parametrized, synthetic stimuli.
One way to assess the quality of an encoding model is to inspect the
tuning functions that are derived from it [\citet{Kay2008}].

For vision, the most fundamental and important kind of tuning function
is the spatial receptive field. Each neuron (or voxel) in each visual
area is sensitive to stimulus energy presented in a limited region of
visual space, and spatial receptive fields describe how the response of
the neuron or voxel is modulated over this region. In the primary
visual cortex, response modulation is typically strongest at the center
of the receptive field. Response modulation is much weaker at the
periphery, but has been shown to have functionally significant effects
on the output of the neuron (or voxel) [\citet{Vinje2000}].

\begin{figure}[b]

\includegraphics{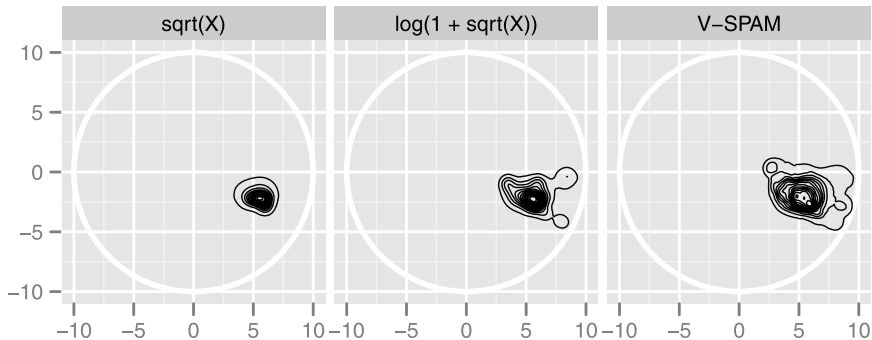}

\caption{Estimated spatial receptive field for voxel 717. The contours
show the predicted response to a point stimulus placed at various
locations across the field of view. They indicate the sensitivity of the
voxel to different spatial locations.}
\label{figspatial-rf}
\end{figure}
%
\begin{figure}

\includegraphics{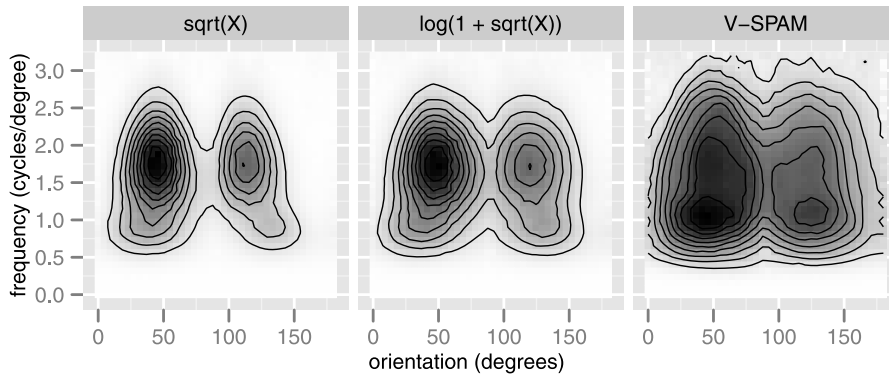}

\caption{Estimated frequency and orientation tuning for voxel 717.
The contours show the predicted response to a 2D cosine stimulus
(a 2D Fourier basis function) parameterized by frequency and orientation.
Darker regions correspond to greater predicted responses. The plot reveals
sensitivity of the voxel to different spectral components.}
\label{figfrequency-and-orientation-tuning}
\end{figure}

The panels in Figure~\ref{figspatial-rf} show estimated spatial
receptive fields for voxel 717 using the three different models
considered here [we chose this voxel because its predictive $R^2$
varied greatly among the three models: 0.26 for the $\mathit{sqrt}(X)$ model
\eqref{eqsqrt-model}, 0.42 for the $\mathit{log}(1+\mathit{sqrt}(X))$ \eqref{eqlog-model}, and 0.57 for V-SPAM \eqref{eqvspam}]. These estimated
receptive fields indicate the locations within the spatial field of
view that are predicted to modulate the response of the voxel by each
model. All three models agree that the voxel is tuned to a region in
the lower-right quadrant of the field of view; however, for V-SPAM the
receptive field is more expansive, and is thus able to capture the weak
but potentially important responses at the far periphery of the visual field.

Like spatial tuning, orientation and frequency tuning are fundamental
properties of V1, so it is essential to inspect the orientation and
frequency tuning functions that are derived from encoding models for
this area. As seen in the panels of Figure~\ref{figfrequency-and-orientation-tuning}, the V-SPAM model is better able
to capture the weaker responses to orientations and spatial frequencies
away from the peaks of the tuning.

\begin{figure}

\includegraphics{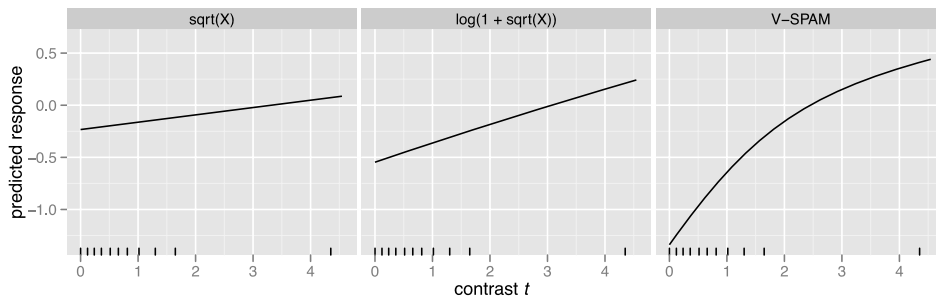}

\caption{Estimated contrast tuning function for voxel 717.
This is the predicted response to a pink noise stimulus at different
levels of RMS contrast $t$. The tick marks indicate the deciles of RMS
contrast in the training images (e.g., fewer than $10\%$ of training
images have contrast between $2$ and~$4$).}
\label{figcontrast-tuning-function}
\end{figure}

Finally, we examine tuning to image contrast, which is another critical
property of V1. Image contrast strongly modulates responses in V1 and
is also perceptually salient, so contrast tuning functions are
frequently used to study the relationship between activity and
perception [\citet{Olman2004}].
The contrast tuning function describes how a voxel is predicted to
respond to different contrast levels. It is constructed by computing
the predicted response to a stimulus of the form $t \cdot w$, where $w$
is standardized 2D pink noise (whose power spectral density is of the
form $1/|\omega|$), and $t \geq0$ is the root-mean-square (RMS)
contrast. At zero contrast the noise is invisible and only the
background can be seen; as contrast increases the noise becomes more
visible and distinguishable from the background. Figure~\ref{figcontrast-tuning-function} shows the contrast response function for
the voxel as estimated by the three models. The first two, the $\mathit{sqrt}(X)$
and $\mathit{log}(1+\mathit{sqrt}(X))$, look nearly linear and relatively flat over the
range of contrasts present in the training images. The V-SPAM
prediction tapers off as contrast increases, and it is much more
negative for low contrasts than predicted by $\mathit{sqrt}(X)$ and
$\mathit{log}(1+\mathit{sqrt}(X))$. The V-SPAM prediction is closer to what is expected
based on previous direct measurements [\citet{Olman2004}], and suggests
that V-SPAM is more sensitive to responses evoked by lower contrast
stimulus energy.

\begin{figure}

\includegraphics{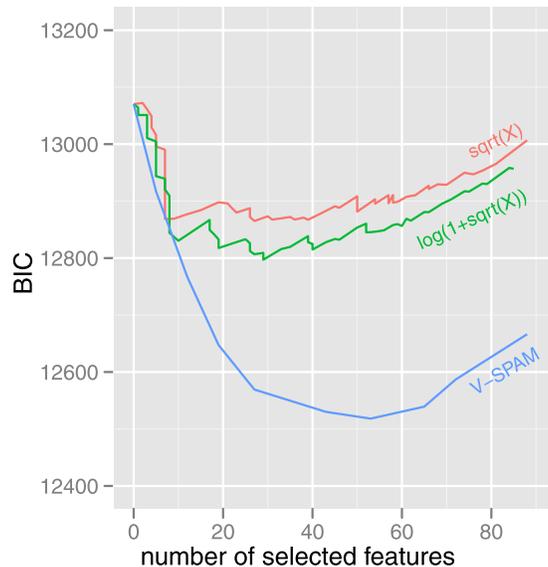}

\caption{Comparison of BIC paths for different models of voxel 717: the $\mathit{sqrt}(X)$
model \protect\eqref{eqsqrt-model}, the $\mathit{log}(1+\mathit{sqrt}(X))$ model
\protect\eqref{eqlog-model}, and V-SPAM \protect\eqref{eqvspam}.}
\label{figbic-comparison}
\end{figure}

The relatively more sensitive tuning functions derived from the V-SPAM
model of voxel 717 have a simple explanation. The models selected by
BIC for this
voxel included different numbers of features: 7 for $\mathit{sqrt}(X)$, 29 for
$\mathit{log}(1+\mathit{sqrt}(X))$, and 53 for V-SPAM. Since the features are localized in
space, frequency, and orientation, the number of features in the selected
model is related to the sensitivity of the estimated tuning functions in
the periphery. BIC forces a trade-off between the residual sum of squares
(RSS) and number of features. The models with fixed transformations
have much
larger RSS values than V-SPAM, and the trade-off (see
Figure~\ref{figbic-comparison}) favors fewer features for them because the
residual nonlinearity (as shown in Figure~\ref{figlinear-voxel-residuals}) does
not go away with increased numbers of features. This suggests that the
sensitivity of a voxel to weaker stimulus energy is not detected by the
$\mathit{sqrt}(X)$ and $\mathit{log}(1+\mathit{sqrt}(X))$ models, because it is masked by residual
nonlinearity. So the tuning function of a voxel can be much broader than
inferred by the model when the model is incorrect.

\section{Conclusion}
\label{secconclusion}

Using residual analysis and a start-of-the-art sparse additive nonparametric
method (SPAM), we have derived V-SPAM encoding models for V1 fMRI BOLD
responses to natural images and demonstrated the presence of an important
nonlinearity in V1 fMRI response that has not been accounted for by previous
models based on fixed parametric nonlinear transforms. This
nonlinearity could
be caused by several different mechanisms including the dynamics of
blood flow
and oxygenation in the brain and the underlying neural processes. By comparing
V-SPAM models with the previous models, we showed that V-SPAM models
can both
improve substantially prediction accuracy for encoding and decrease
substantially identification error when decoding from very large collections
of images. We also showed that the deficiency of the previous encoding
models with fixed parametric nonlinear transformations also affects tuning
functions derived from the fitted models.

Since encoding and decoding models are becoming more prevalent in fMRI
studies, it is important to have methods to adequately characterize the
nonlinear aspects of the response-stimulus relationship. Failure to address
nonlinearity effectively can lead to suboptimal predictions and
incorrect inferences. The methods used here, combining residual
analysis and sparse nonparametric modeling, can easily be adopted by
neuroscientists studying any part of the brain with encoding and
decoding models.

\begin{appendix}\label{appm}

\section*{Appendix: Extracting the fMRI BOLD response}
\label{sechemodynamic}

\begin{figure}

\includegraphics{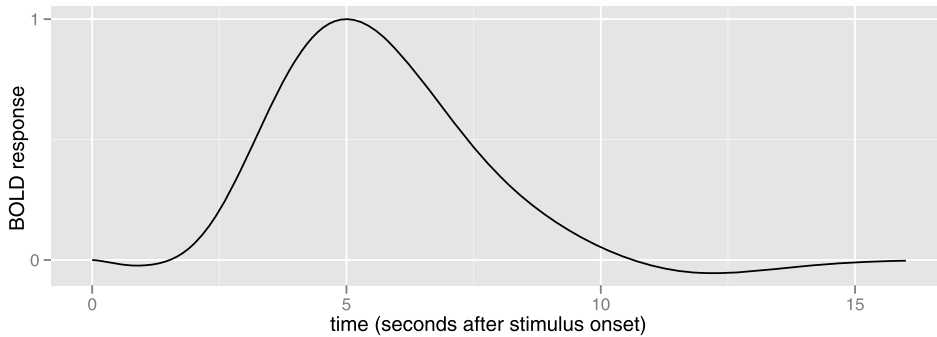}

\caption{A model hemodynamic response function.%
}
\label{fighrf}
\end{figure}

The {fMRI} signal $Z_v(t)$ measured at voxel $v$ can be modeled as a
sum of
three components: the BOLD signal $B_v(t)$, a nuisance signal $N_v(t)$
(consisting of low frequency fluctuations due to scanner drift, physiological
noise, and other nuisances), and noise $\epsilon_v(t)$:
\[
Z_v(t)
= B_v(t) + N_v(t) + \epsilon_v(t)
 .
\]
The BOLD signal is a mixture of evoked responses to image stimuli.
This reflects the underlying hemodynamic response that results from neuronal
and vascular changes triggered by an image presentation. The hemodynamic
response function $h_v(t)$ characterizes the shape of the BOLD response (see
Figure~\ref{fighrf}), and is related to the BOLD signal by the linear time
invariant system model [\citet{Friston1994}],
\[
B_v(t)
= \sum_{k=1}^n \sum_{\tau\in T_k} A_v(k) h_v(t - \tau)
 ,
\]
where $n$ is the number of images, $T_k$ is the set of times at which
image $k$ is
presented to the subject, and $A_v(k)$ is the amplitude of the voxel's
response to image~$k$.

To extract $A_v(\cdot)$ from the {fMRI} signal, it is necessary to
estimate the
hemodynamic response function and the nuisance signal. We used the
method described in \citet{Kay2008hbm}, modeling $h_v(t)$ as a linear
combination of Fourier basis functions covering a period of 16 seconds
following stimulus onset, $N_v(t)$ as a degree 3 polynomial, and
$\epsilon_v(t)$ as a first-order autoregressive process. The resulting
estimates $\hat{A}_v(\cdot)$ are the voxel responses for each image.
\end{appendix}

\section*{Acknowledgments}
\label{secacknowledgments}
 A preliminary version of this work was presented in \citet
{VSPAM2009}. We thank the Editor and reviewer for valuable comments on
an earlier version that have led to a much improved article.

%
%



\printaddresses

\end{document}